\documentstyle[12pt]{article}

\begin{document}

\def\ve#1{\mid #1\rangle}
\def\vc#1{\langle #1\mid}

\newcommand{\p}[1]{(\ref{#1})}
\newcommand{\be}{\begin{equation}}
\newcommand{\ee}{\end{equation}}

\newcommand{\sect}[1]{\setcounter{equation}{0}\section{#1}}
\renewcommand{\theequation}{\thesection.\arabic{equation}}

\newcommand{\vs}[1]{\rule[- #1 mm]{0mm}{#1 mm}}
\newcommand{\hs}[1]{\hspace{#1mm}}
\newcommand{\mb}[1]{\hs{5}\mbox{#1}\hs{5}}
\newcommand{\Db}{{\overline D}}
\newcommand{\bea}{\begin{eqnarray}}
\newcommand{\eea}{\end{eqnarray}}
\newcommand{\wt}[1]{\widetilde{#1}}
\newcommand{\und}[1]{\underline{#1}}
\newcommand{\ov}[1]{\overline{#1}}
\newcommand{\sm}[2]{\frac{\mbox{\footnotesize #1}\vs{-2}}
           {\vs{-2}\mbox{\footnotesize #2}}}
\newcommand{\prt}{\partial}
\newcommand{\eps}{\epsilon}

\newcommand{\R}{\mbox{\rule{0.2mm}{2.8mm}\hspace{-1.5mm} R}}
\newcommand{\Z}{Z\hspace{-2mm}Z}

\newcommand{\cd}{{\cal D}}
\newcommand{\cg}{{\cal G}}
\newcommand{\ck}{{\cal K}}
\newcommand{\cw}{{\cal W}}

\newcommand{\vj}{\vec{J}}
\newcommand{\vl}{\vec{\lambda}}
\newcommand{\vz}{\vec{\sigma}}
\newcommand{\vt}{\vec{\tau}}
\newcommand{\vw}{\vec{W}}
\newcommand{\poiss}{\stackrel{\otimes}{,}}

\def\l#1#2{\raisebox{.2ex}{$\displaystyle
  \mathop{#1}^{{\scriptstyle #2}\rightarrow}$}}
\def\r#1#2{\raisebox{.2ex}{$\displaystyle
 \mathop{#1}^{\leftarrow {\scriptstyle #2}}$}}

\title{New approach to representation theory of semisimple Lie algebras
and quantum algebras}
\author{A.N. Leznov$^{a,b,}$\footnote{E-mail: leznov@ce.ifisicam.unam.mx}
\\
{\small \em {~$~^{(a)}$ Institute for High Energy Physics,}}\\
{\small \em 142284 Protvino, Moscow Region, Russia}\\
{\small \em {~$~^{(b)}$ Bogoliubov Laboratory of Theoretical Physics,
JINR,}}\\
{\small \em 141980 Dubna, Moscow Region, Russia}}
\date{}
\maketitle

\begin{abstract}
A method to construct in explicit form the generators of the simple roots 
of an arbitrary finite-dimensional representation of a quantum or standard 
semisimple algebra is found. The method is based on general results from
the 
global theory of representations of semisimple groups. The rank two
algebras 
$A_2$, $B_2=C_2$, $D_2$ and $G_2$ are considered as examples. 
The generators of the simple roots are presented as solutions of a system
of 
finite difference equations and given in the form of  $N_l\times N_l$
matrices, 
where $N_l$ is the dimension of the representation.
\end{abstract}

\sect{Introduction}

The basis  vectors of a finite-dimensional irreducible representation 
$l=(l_1, l_2, \ldots, l_r)$ of a semisimple algebra are usually constructed
by 
repeated application of the lowering  generators $X^-_i$ to the highest
vector 
$\ve{l}$ with the properties:
$$
X^+_s \ve{l}=0,\quad h_s \ve{l}=l_s \ve{l}
$$
where $X^{\pm}_s$, $h_s$ are respectively the generators corresponding to
the 
simple roots and the Cartan elements.
The obvious problem with such a construction is that not
all state vectors  arising in this fashion are linearly independent and an 
additional procedure for excluding linearly dependent components with
further 
orthogonalization of the basis is necessary. Usually this is not a simple 
matter.
Nevertheless, the values which the group element 
$\exp \tau\equiv \exp \sum h_i\tau_i$ 
takes on basis vectors may be obtained from the invariant Weyl 
character formula for the irreducible representation $l=\sum h_i l_i$,
\begin{equation}
\pi^l(\exp \tau)={ \sum_W \delta_W \exp (\tau_W, l+{1\over 2} \rho)\over 
\sum_W \delta_W \exp (\tau_W, {1\over 2} \rho)}\ .\label{WL}
\end{equation}
Presented as a sum of exponents, (the denominator is always a divisor of
the 
numerator!):
$$
\pi^l(\exp \tau)=\sum_{n^k}^{N_l} C_{n^k}\exp (\tau, n^k) =\sum_{n^k}^{N_l} 
C_{n^k}\exp \sum_i^r (\tau_i n^k_i)\ ,
$$
this gives answers to many questions about the structure of the basis of
the 
corresponding representation. In (\ref{WL}) $W$ is the element of the
discrete 
Weyl group, $\delta_W$ is its signature, $\tau_W$ is the result of the
action of
the group element $W$ on $\tau$, $C_{n^k}$ is the multiplicity of
corresponding 
exponent and  $ \rho$ is the  sum of the positive 
roots of the corresponding algebra. 

Recalling the definition,
$$
\pi^l(\exp \tau)= Trace (\exp \tau)=\sum_{\alpha}^{N_l} \vc{\alpha} (\exp
\tau)
\ve{\alpha},
$$
where $\vc{\alpha},\ve{\alpha}$ are the bra and ket basis vectors of 
the representation $l$, we see that the Weyl formula yields the action of
the 
group element $\exp \tau$ on basis vectors with a given number of lowering 
operators of various types, namely,
\begin{equation}
e^\tau (X^-_1)^{m_1}...(X^-_r)^{m_r} \ve{l}= 
e^{ \sum \tau_i l_i} e^{-\sum_{p,s} m_p\tilde K_{p,s} \tau_s }
(X^-_1)^{m_1}...(X^-_r)^{m_r}\ve{l}
\label{L} 
\end{equation}
(of course the ordering of the lowering operators is inessential in this 
expression).
Equating each exponent of the Weyl formula to the corresponding exponent of 
(\ref{L}) we easily find the indices $m_i$. The only thing 
the Weyl formula cannot do is to distinguish among the basis vectors
arising from the same set of lowering operators  but taken in a 
different order. Nevertheless,  the number of such states is given by the
multiplicity of the corresponding exponent in the character formula.

In the present paper the above comments play a key role. We present
an alternative way to construct  in explicit form the generators of the 
simple roots. Instead of the procedure of excluding linearly dependent 
components and subsequent orthogonalization, we need to solve a system 
of finite difference equations whose solvability 
is guaranteed by the global theory of representations of 
semisimple algebras.
In principle, these conditions can be found independently (as conditions
for
the solvability of such system with  "fixed" boundary), but they may be
seen to be equivalent to  known results in  global representation theory.

Briefly, our proposed program is as follows. The initial data
are the known dimensions and characters of irreducible representations of 
semisimple groups, given by the  famous Weyl formulae \cite{Weyl}.  
The final ones are the explicit forms of the generators of 
the simple roots of both quantum and usual semisimple algebras. 
We do not distinguish between  bases for quantum
and standard semisimple algebras; and present in  
explicit form the generators of the simple roots in the form of 
$(N_l\times N_l)$ matrices ($N_l$ is the dimension of the corresponding
representation given by the Weyl dimension formula), passing over the
question 
about the structure of the basis.
 
The paper is organized in the following way. In section 2 we rewrite
defining 
relations of a quantum algebra in terms of only $2r$ generators instead of 
the $3r$ generators of the traditional approach. 
This construction is concretised for algebras of 
second rank in section 3. In three subsequent sections, 4, 5, and 6, the 
calculations for the algebras $A_2$, $B_2=C_2$ and $G_2$ are presented 
in detail. Concluding remarks and perspectives for further investigation
are 
gathered in section 7.

\sect{Modified form of equations quantum algebras defined} 

The customary form of $3r^2$ commutation relations among the 
$3r$ generators, the 
simple roots $X^{\pm}_i$  and Cartan elements $h_i$, of quantum algebras 
is given by:
\begin{equation}
[h_i,X^{\pm}_j]=\pm K_{j,i} X^{\pm}_j,\quad [X^+_i,X^-_j]=\delta_{j,i}
{\sinh (t w_i h_i)\over \sinh (w_i t)}\label{1}
\end{equation}
where  $K$ is the Cartan matrix, $K_{j,i}w_i=w_jK_{j,i}\equiv \tilde
K_{j,i}$,
$t$ is the deformation parameter. The Cartan matrices of the series
$A_n,D_n,
E_{6,7,8}$ are a priori symmetric with $\tilde K_{i,j}= K_{i,j}\,,\ w_i=1$. 

Let us introduce the alternative set of $3r$ generators belonging to the
universal enveloping algebra,
$$
T^{\pm}_i= e^{w_i t h_i\over 4} X^{\pm}_i e^{w_i th_i\over 4},\quad 
R_i= e^{ t w_i h_i}
$$
In terms of these the relations determining quantum
algebra (\ref{1}) may be rewritten,
\begin{equation}
R_iT^{\pm}_j= e^{ \pm \tilde K_{j,i}t} T^{\pm}_jR_i,\quad 
e^{\tilde K_{j,i} t \over 2}T^+_iT^-_j-e^{-\tilde K_{j,i}t\over 2}  T^-_j
T^+_i=\delta_{i,j}{R_i^2-1\over 2\sinh w_i t}\ .\label{2}
\end{equation}
Introducing $2r$ generators,
$$
Q^{\pm}_i=T^{\pm}_i\pm {R_i\over 2\sinh w_it}
$$
the system (\ref{2}) takes the form
$$
e^{ w_i t} Q^+_iQ^-_i- e^{-w_i t} Q^-_iQ^+_i=-{1\over 2\sinh w_it}
$$
$$
e^{\tilde K_{j,i}t\over 2}  (Q^+_iQ^-_j-Q^+_jQ^-_i)-
e^{-\tilde K_{j,i}t\over 2}  (Q^-_jQ^+_i-Q^-_iQ^+_j)=0,\quad \tilde K_{i,j}
\neq 0
$$
\begin{equation}
[Q^+_i,Q^-_j]=[Q^+_j,Q^-_i]=0,\quad \tilde  K_{i,j}=0\label{3}
\end{equation}
\begin{eqnarray*}
&& e^{\tilde K_{j,i}t\over 2}  (Q^+_iQ^-_j+Q^+_jQ^-_i) 
- e^{-\tilde K_{j,i}t\over 2}  (Q^-_jQ^+_i+Q^-_iQ^+_j) \\
&& \hskip 8em {} = -{\sinh {\tilde K_{j,i}\over 2} t\over
\sinh w_i t \sinh w_j t} R_i R_j,\quad \tilde K_{i,j}\neq 0,\quad i\neq j
\end{eqnarray*}
$$
R_iQ^{\pm}_j = 
e^{ \pm \tilde K_{j,i}t} Q^{\pm}_jR_i\mp (e^{ \pm \tilde K_{j,i}t}-1) 
{R_jR_i\over 2\sinh w_j t}
$$
We note that the first three rows relate the  $Q^{\pm}_i$
among themselves. It is therefore possible to
consider these relations as some subalgebra of the 
universal enveloping  algebra (\ref{1}).

Now we would like to show that as a direct corollary of equations (\ref{3})
all generators $R_i$ may be expressed algebraically as functionals of 
generators $Q^{\pm}_s$. For this purpose let us multiply the equation of
the 
last row of (\ref{3}) (with exchange indices $i\to j$) on $R_i$ from the
left. 
We obtain
$$
R_iR_jQ^{\pm}_i= 
e^{ \pm \tilde K_{i,j} t} R_iQ^{\pm}_iR_j \mp\ (e^{\pm \tilde K_{ij}t}-1)
{R^2_iR_j\over 2\sinh w_it}=
$$
\begin{equation}
e^{\pm(\tilde K_{ii}+\tilde K_{i,j} t})Q^{\pm}_iR_iR_j\ 
\mp\ (e^{\pm(\tilde K_{ii}+ \tilde K_{ij} t)}-1)
{R_iR_j R_i\over 2\sinh w_i t}
\label{4}
\end{equation}
Introducing the operator 
$$
Q_{i,j}\equiv e^{\tilde K_{ij} t\over 2}(Q^+_iQ^-_j+Q^+_jQ^-_i)-
e^{-{\tilde K_{ij}t\over 2}}(Q^-_jQ^+_i+Q^-_iQ^+_j)=Q_{j,i}\ ,
$$
which in virtue of (\ref{3}) is proportional to $R_iR_j$, we rewrite 
(\ref{4}) in the form,
\begin{equation}
Q_{ij}Q^{\pm}_iQ^{-1}_{ij}= e^{\pm(\tilde K_{ii}+\tilde K_{ij})t}
Q^{\pm}_i\mp
\left(e^{\pm ( \tilde K_{ii}+\tilde K_{ij})t}-1\right){R_i\over 2\sinh
w_it}\ .
\label{5}
\end{equation}
{}From (\ref{5}) we conclude that in the case 
$\tilde K_{ii}+\tilde K_{ij} \neq 0$ the proposition  above is true
and the generator $R_i$ may be expressed algebraically in terms of the
generators $Q^{\pm}$.
In the case $\tilde K_{ii}+\tilde K_{ij}=0$, definitely 
$\tilde K_{jj}+ \tilde K_{ij}\neq 0$ and generator $R_j$ can be expressed 
in terms of $Q$ generators. 
Then $R_i$ can be found from the equation relating $Q_{i,j}$ to the product 
$R_iR_j$. Thus the above assertion is true in all cases.

In the next section we specify ourselves to the rank two cases. 
The rank one case is well known and we summarize it for later use.
Irreducible representations of the $A^q_1$ algebra are labelled by natural 
or half natural number $l$, with dimension of the representation given by
$2l+1$. The generator $H$ takes all odd or even values between $2l$ and
$-2l$;  $H_k=2l-2k\equiv 2m$, $0\leq k \leq 2l$. The non-zero matrix 
elements of generators $X^{\pm}$ are 
$$
X^{\pm}_{m,m\pm 1}\ =\
\left({\sinh(l\mp m)t\over \sinh t} 
{\sinh(l\pm m+1)t\over \sinh t}\right)^{1\over 2}
$$
where $\,m{=}l{-}k\,$ and the condition $\,X^+=(X^-)^T\,$ is satisfied.
The structure of the $Q^{\pm}$ generators is as follows: diagonal elements 
are 
$Q^{\pm}_{m,m}= {e^{\pm 2(l-k)t}\over 2\sinh t}$; and the non-zero 
nondiagonal elements,
$$
Q^{\pm}_{m,m\pm 1}=e^{(m\pm {1\over 2})t}
\left({\sinh(l\mp m)t\over \sinh t} 
{\sinh(l\pm m+1)t\over \sinh t}\right)^{1\over 2}\ .
$$

\sect{The algebras of the second rank}

In this section we restrict ourselves indices taking only two values 
$i=1,2$ in the general
system (\ref{3}) or to algebras of  second rank, $A_2,B_2=C_2,G_2$.
The symmetrical Cartan matrices for these algebras have the form,
$$
\tilde K=\pmatrix{ 2 & -p \cr
                  -p & 2p \cr},\quad w_1=1,\quad w_2=p\ ,
$$
where $\,p=1,2,3\,$ for the cases $A_2,B_2,G_2$ respectively.
The following additional abbreviations are suitable:
$$
Q^{\pm}_1={s^1\pm r^1\over 2\sinh t},\quad Q^{\pm}_2={s^2\pm r^2\over
2\sinh pt}
$$                   
The first two rows of (\ref{3}) take the form:
\begin{equation}\begin{array}{rll}
[s^1,r^1] &=& \tanh t\left((s^1)^2-(r^1)^2+1\right)\\[5pt] 
[s^2,r^2] &=& \tanh pt((s^2)^2-(r^2)^2+1)\\[5pt]
[s^1,s^2]-[r^1,r^2] &=& \tanh {pt\over 2}(\{r^1,s^2\}-\{s^1,r^2\})\label{6}
\end{array}\end{equation}
The  operator $Q_{1,2}\equiv Q$ is given by
(we use a rescaled version of the general definition in the previous
section),
\begin{equation}
Q\ =\ {\sinh t\sinh pt\over \sinh {pt\over 2}}
\left(e^{-pt\over 2}(Q^+_1Q^-_2+Q^+_2 Q^-_1\right) Q-
e^{pt\over 2}(Q^-_1Q^+_2+Q^-_2Q^+_1) 
\ =\ R_1R_2 
\label{6A}
\end{equation}
The operators $R_1,R_2$ are expressed in terms of operators
$Q,Q^{\pm}_{1,2}$ 
in virtue of the relations
$$
QQ^{\pm}_1Q^{-1}\ =\ 
e^{ \mp(p-2)t} Q^{\pm}_1\ \mp\ \left(e^{\mp(p-2)t}-1\right){R_1\over 2\sinh
t},
$$
\begin{equation}
QQ^{\pm}_2Q^{-1}\ =\ 
e^{ \pm pt} Q^{\pm}_2\ \mp\ \left(e^{\pm pt}-1\right){R_2\over \sinh pt}
\label{7}
\end{equation}
In terms of $s^{1,2},r^{1,2}$ these relations may be rewritten in form more 
suitable for our purposes:
$$
\sinh (p-2)t R_1=\sinh (p-2)t r^1+Qs^1Q^{-1}-\cosh (p-2)t s^1
$$
\begin{equation}
-\sinh pt R_2=-\sinh pt r^2+Qs^2Q^{-1}-\cosh pt s^2\ .\label{7A}
\end{equation}
We note that in the case of the $B_2=C_2$ algebra ($p=2$) operator $R_1$
cannot be defined from (\ref{7}). But $R_2$ is well defined and
$R_1$ may be algebraically expressed after this from the equation for $Q$ 
operator. Eliminating $R_{1,2}$ from equations (\ref{7}) we conclude that,
$$
Q \left( e^{{(p-2)t\over 2}} Q^+_1-e^{-{(p-2)t\over 2}} Q^-_1 \right)
Q^{-1}
\ =\ {(p-2)t\over 2} Q^+_1-e^{{(p-2)t\over 2}} Q^-_1
$$
$$
Q\left(e^{-{pt\over 2}} Q^+_2 - e^{{pt\over 2}} Q^-_2\right)Q^{-1}\ =\
e^{{pt\over 2}} Q^+_2 - e^{-{pt\over 2}} Q^-_2\ .
$$
These equations are equivalent to
$$
\cosh{(p-2)t\over 2} \left(r^1-Qr^1Q^{-1}\right)=\sinh{(p-2)t\over 2}
(s^1+Qs^1Q^{-1}),
$$
\begin{equation}
\cosh{pt\over 2} \left( r^2 - Qr^2Q^{-1}\right)\ =\ 
 -\sinh{pt\over 2} \left( s^2 + Qs^2Q^{-1}\right)
\label{8}
\end{equation}
Using the fact that operators $R_1,R_2,Q$ are mutual commutative as a
direct consequence of (\ref{7}) we obtain the following important relations
\begin{equation}
Qs^1Q^{-1} + Q^{-1}s^1Q\ =\  2\cosh(p-2) t s^1,\quad
Qs^2Q^{-1} + Q^{-1}s^2Q\ =\  2\cosh p t s^2\label{9}
\end{equation}
{}From these relations we conclude that in representations with diagonal
$Q$ operator, matrix elements of generators $s^{1,2}$ satisfy the equations
\begin{equation}
\left( {\lambda_i\over \lambda_j} + {\lambda_j\over \lambda_i}
 - e^{(p-2)t} - e^{-(p-2)t}\right) s^1_{i,j}\ =\ 0\ , \quad 
\left({\lambda_i\over \lambda_j}+{\lambda_j\over \lambda_i} 
  - e^{pt} - e^{-pt} \right) s^2_{i,j}\ =\ 0\ .\label{10}
\end{equation}
In other words the matrix elements $s^1_{i,j}$ are different from zero only
in the case when ${\lambda_i\over \lambda_j}=e^{\pm (p-2)t}$, 
and similarly, the  matrix elements $s^2_{i,j}$ are nonzero when
${\lambda_i\over \lambda_j}=e^{\pm pt}$.

In the next sections, we will concretize these relations for the three
individual cases of $A_2, B_2=C_2, G_2$ algebras.

\section{The $A_2$ algebra case}

We work in the basis with  diagonal $Q$ generator. From (\ref{10}) ($p=1$)
we can conclude that 
${\lambda_i\over \lambda_{i+1}}=e^ t$ (keeping in mind irreducibility
of $s^1,r^1$ matrices). But about  multiplicity of 
each $\lambda_i$  nothing can be deduced and we assume it to 
be  arbitrary, $N_i$. 
{}From (\ref{9}) follows the "row" structure of the $s^{1,2}$ generators.
Namely
$$
s^1=\pmatrix{....& a_{i,i-1} & 0 & a_{i,i+1}... \cr},\quad
s^2=\pmatrix{....& b_{i,i-1} & 0 & b_{i,i+1}... \cr}
$$
where $a_{i,i-1}, b_{i,i-1}$ are rectangular matrices of the dimension
$N_{i-1}\times N_i$ wether as $a_{i,i+1}, b_{i,i+1}$ are rectangular
matrices 
of the dimensions $N_i\times N_{i+1}$ ( zero in the center is quadratic
$N_i\times N_i$ zero matrix).
{}From equations (\ref{8}) ($p=1!$) it is possible to reconstruct the "row"
structure of $r^{1,2}$ matrices, namely, 
$$
r^1=\pmatrix{....& -a_{i,i-1} & \alpha_i & a_{i,i+1}... \cr},\quad
r^2=\pmatrix{....& -b_{i,i-1} & \beta_i & b_{i,i+1}... \cr}
$$
where $\alpha_i,\beta_i$ are quadratic $N_i\times N_i$ matrices.
The system of equations which matrices $a,b,\alpha, \beta$ satisfy arises
after substitution of these ans\"atze for $s,r$ into (\ref{6}) and
(\ref{6A}).
Equation relating $s^1,r^1$ is equivalent to a matrix system:
$$
e^{-t} a_{n,n-1} \alpha_{n-1}=e^t \alpha_n a_{n,n-1},\quad 
e^t a_{n,n+1} \alpha_{n+1}=e^{-t} \alpha_n a_{n,n+1}, 
$$
\begin{equation}
2e^{-t} a_{n,n-1}a_{n-1,n}-2e^t a_{n,n+1}a_{n+1,n}=\sinh t(I_n-\alpha_n^2)
\label{11}
\end{equation}
The same for $s^2,r^2$ leads to:
$$
e^{-t} b_{n,n-1} \beta_{n-1}=e^t \beta_n b_{n,n-1},\quad 
e^t b_{n,n+1} \beta_{n+1}=e^{-t} \beta_n b_{n,n+1}, 
$$
\begin{equation}
2e^{-t} b_{n,n-1} b_{n-1,n}-2e^t b_{n,n+1} b_{n+1,n}=\sinh t(I_n-\beta_n^2)
\label{12}
\end{equation}
where $I_n$ denotes the $N_n\times N_n$  unit matrix.
The last equation in (\ref{6}) and the definition of $Q$ (\ref{6A})
imply the system,
$$
e^{{t\over 2}} b_{n,n-1}\alpha_{n-1}=e^{-{t\over 2}}\alpha_n
b_{n,n-1},\quad 
e^{-{t\over 2}} a_{n,n+1}\beta_{n+1}=e^{{t\over 2}}\beta_n a_{n,n+1}, 
$$
$$
e^{{t\over 2}} a_{n,n-1}\beta_{n-1}=e^{-{t\over 2}}\beta_n a_{n,n-1},\quad 
e^{-{t\over 2}} b_{n,n+1}\beta_{n+1}=e^{{t\over 2}}\alpha_n b_{n,n+1}, 
$$
\begin{equation}
2e^{-{t\over 2}}\left({\alpha_n \beta_n\over 4} 
+ b_{n,n+1}a_{n+1,n}\right)-2e^{{t\over 2}}
\left({\beta_n\alpha_n\over 4}+ a_{n,n-1}b_{n-1,n}\right)
=\sinh {t\over 2} \lambda_n I_n 
\label{13}
\end{equation}
At first sight the system (\ref{11}),(\ref{12}) and (\ref{13}) is so 
complicated that all attempts to solve it seem to have a little chance for 
success. But this is not so, as we shall demonstrate in the next
few pages.
First of all let us substitute the ansatz for $s^1,r^1,s^2,r^2$ into
(\ref{7A})
($p=1$). Direct calculations show that $R_1,R_2$ are the 
block-diagonal matrices with block matrices $\alpha_n,\beta_n$,
respectively.
Moreover the mutual commutativity of $R_1,R_2$ have as its consequence the
mutual commutativity  $[\alpha_n,\beta_n]=0$. Equation (\ref{6A}) relating
$Q$ with $R_1R_2$ may be rewritten in the form: 
$$
\lambda_n I_n=\alpha_n \beta_n.
$$
Taking into account these circumstances we can eliminate matrices $\beta_n$
from the system (\ref{11})-(\ref{13}) and rewrite it in the form,
\begin{equation}
\begin{array}{rll}
&& e^{-t} a_{n,n-1} \alpha_{n-1}=e^t \alpha_n a_{n,n-1}
\quad,\quad 
   e^t a_{n,n+1} \alpha_{n+1}=e^{-t} \alpha_n a_{n,n+1}\\[5pt] 
&& e^{{t\over 2}} b_{n,n-1}\alpha_{n-1}=e^{-{t\over 2}}\alpha_n b_{n,n-1}
\quad,\quad 
e^{-{t\over 2}} b_{n,n+1}\alpha_{n+1}=e^{{t\over 2}}\alpha_n
b_{n,n+1},\\[5pt] 
&&
2e^{-t} a_{n,n-1}a_{n-1,n}-2e^t a_{n,n+1}a_{n+1,n}
= \sinh t(I_n-\alpha_n^2) \\[5pt]
&& 2e^{-t} b_{n,n-1} b_{n-1,n}-2e^t b_{n,n+1} b_{n+1,n}
= \sinh t(I_n-\lambda_n^2 \alpha_n^{-2}) \\[5pt]
&&
e^{-{t\over 2}} b_{n,n+1}a_{n+1,n}-e^{{t\over 2}} a_{n,n-1}b_{n-1,n}=0
\label{14}
\end{array}
\end{equation}
As  mentioned above the matrix $\alpha_n$ commutes with $\beta_n$ and
so they both simultaneously can be presented in diagonal form. The diagonal 
elements of $\,\alpha_n\,,\,\beta_n\,$ we will denote by double indices
$\alpha^s_n\,,\,\beta^s_n\,,\ 1\leq s \leq N_n$. 
Obviously $\alpha^s_n \beta^s_n=\lambda_n$.

The system (\ref{14}) in the presented form is unlimited  and for 
its solution some additional "boundary" conditions are necessary. To have  
solution in the form of finite-dimensional matrices we will assume that
on its ``left end" $a_{1,0}=b_{1,0}=0$ and on its ``right end" $a_{N,N+1}=
b_{N,N+1}=0$. The numbers $N_n$ and diagonal elements
$\lambda_n,\alpha_n^s$
must be found as conditions for resolving  (\ref{14}) under such boundary 
conditions. 
We shall use known facts from global representation theory of the 
$A_2$ algebra to resolve the system (\ref{14}) explicitly and thus
obtain explicit forms for the generators of the simple roots for the
irreducible representations $(p,q)$ of the $A_2$ algebra.


As was mentioned in the introduction basis vectors of 
irreducible representations of
semisimple algebras are constructed from the single highest vector
(\ref{1})
by action  of lowering operators. 
{}From the definition of the highest vector (\ref{1}) 
it follows immediately that 
$\lambda_1=e^{(p+q)t}$, $\alpha_1^1=e^{pt}$ and $N_1=1$. We recall that 
$\alpha=\exp h_1t\ ,\ \lambda=\exp (h_1+h_2)t$.
Let us consider the next bases vectors which arise after action by
generators $X^-_{1,2}$ on highest vector. It is obvious that
$$
N_2=2, \quad (q\neq 0), \quad \lambda_2=e^{(p+q-1)t},\quad
\alpha_2^1=e^{(p-2)t}
,\quad \alpha_2^2=e^{(p+1)t}
$$
Let us for the meantime set aside the problem of linearly
independent components.
Then at each  step after application of two generators $X^-_{1,2}$ to
each basis vector of the previous step the number of basis vectors will be
twice
that on the previous step and the following relations become obvious
$$
N_s=2^{s-1},\quad \lambda_s=e^{(p+q-s+1)t},\quad
\alpha_s^k=e^{(p-2(s-1)+3k)t},
\quad (C_{s-1}^k), \quad 0\leq k \leq (s-1)
$$
The multiplicities $C^k_r$ are binomial coefficients.

Now, let us return to a real situation. The character of irreducible 
representation $(p,q)$ of $A_2$ algebra is given by the Weyl formula
\begin{equation}
\pi^{(p,q)}(\exp(\tau_1h_1+\tau_2h_2)=
{Det \pmatrix{e^{\tau_1 l_1} & e^{\tau_1 l_2} & e^{\tau_1 l_3} \cr 
 e^{(\tau_2-\tau_1) l_1} & e^{(\tau_2-\tau_1) l_2} & e^{(\tau_2-\tau_1)
 l_3} \cr
              e^{-\tau_2 l_1} & e^{-\tau_2 l_2} & e^{-\tau_2 l_3} \cr}\over
Det \pmatrix{e^{\tau_l} & 1 & e^{-\tau_1} \cr 
 e^{(\tau_2-\tau_1)} & 1 & e^{-(\tau_2-\tau_1)} \cr
              e^{-\tau_2} & 1 & e^{\tau_2} \cr}}\label{15}
\end{equation}
where $l_1-l_2=p+1,l_2-l_3=q+1,l_1+l_2+l_3=0$.              
Evaluating the determinants leads to,
\bea
\pi &=& \frac{1}{1-e^{-(\tau_2+\tau_1)}} [ e^{(\tau_2-\tau_1)(p-q)}(e^{px}+
\cdots + 1)(e^{qy}+ \cdots +1) \nonumber \\
&& \hskip 4em {} - e^{-\tau_1(q+1)}e^{-\tau_2(p+1)}(e^{qx}+ \cdots
+1)(e^{py}+ \cdots +1)], \label{HR}
\eea
where $x=2\tau_1-\tau_2$, $y=2\tau_2-\tau_1$, $x+y=\tau_2+\tau_1$.
It is obvious that under the condition $\tau_2+\tau_1=0, x+y=0$ the
numerator 
is equal to zero and and thus the last expression passes to the sum of
exponentials the
arguments of which are different linear combinations of $\tau$ with
definite coefficients.

The reduction of (\ref{15}) or (\ref{HR}) to the $A_1$ subgroup with the 
infinitesimal generators $X^+=[X^+_1,X^+_2],X^-=[X^-_1,X^-_2],H=h_1+h_2$ is 
equivalent to the substitution $\tau_1=\tau_2=t$, yielding the final
result:
\bea
\pi^{p,q}(\exp (h_1+h_2)t) &=& e^{-(p+q)t}{\sum^{p+q+1}_{k=0}
e^{(p+q+1-k)t}
\sum^q_{r=0} e^{(q-r)t} \sum^p_{s=0} e^{(p-s)t}\over (e^t+1)}
\nonumber\\
&=& \sum^{2(p+q)}_{m=0} e^{(p+q-m)t} c_m(p,q)
\eea
Among three natural numbers $p,q,p+q+1$ at least one is odd and by this
reason
numerator is always divided on denominator.
The reduction to the $A_1$ subgroups with
the algebras of the first and second simple roots $X^{\pm}_{1,2}, h_{1,2}$
leads to the same expressions exponents in Weyl formula but in the
different 
order. This result is obvious without any calculations, because Cartan
elements 
of all roots of $A_2$ algebra are relating by discrete Weyl
transformations. 
Before the general consideration we would like to consider the 
concrete example of the
$(2,1)$ representation of $A_2$ algebra. This example  will reveal the main
points of the whole construction.

\subsection{$(2,1)$ representation of $A_2$ algebra}

The direct calculation of the character of the 
$(2,1)$ representation leads to the
following sum of exponents ( in connection with the (\ref{HR})), which we
have 
written in a definite order the sense of which will be soon understandable:
$$
\pi^{(2,1)}(\tau)=e^{(2\tau_1+\tau_2)}+e^{2\tau_2}+e^{(3\tau_1-\tau_2)}+
e^{(-2\tau_1+3\tau_2)}+2e^{\tau_1}+2e^{(-\tau_1+\tau_2)}+e^{(-2\tau_1+2\tau
_2)}
$$
$$
e^{-3\tau_1+2\tau_2}+2e^{-\tau_2}+e^{-2\tau_1}+e^{(\tau_1-3\tau_2)}+
e^{-(\tau_1+2\tau_2)}
$$
Let us now consider reducing of this expression to the $SL(2,R)$ subgroup
related to the composite root of $A_2$. Writing $\tau_1=\tau_2=t$ 
we obtain,
$$
\pi=e^{3t}+2e^{2t}+3e^{t}+3+3e^{-t}+2e^{-2t}+e^{-3t}
$$
Since the generator $h_1+h_2$ takes unit values on each simple root of the
$A_2$ algebra, it follows that the states of highest (lowest) vectors 
are the singlet ones, subspaces with two or five lowering operators are
two-
dimensional and so on. The values taken by operator $R_1R_2\equiv \exp 
(h_1+h_2)t$ on the basis vectors of the $(2,1)$ representation are
precisely 
the exponents of the sum above taken in the same order with the same 
multiplicity.  
By the arguments of the same kind we find the values, which generators
$R_1,R_2$
take on this basis. We present them in the row form performing the correct 
order: 
$$
R_1=(e^{2t},1,e^{3t},e^{-2t},e^t,e^t,e^{-t},e^{-t},e^{2t},e^{-3t},1,1,e^{-2
t},
e^t,e^{-t})
$$
$$
R_2=(e^t,e^{2t},e^{-t},e^{3t},1,1,e^t,e^t,e^{-2t},e^{2t},e^{-t},e^{-t},1,
e^{-3t},e^{-2t})
$$
These define simultaneously the dimensions 
of $\alpha,\beta$ matrices and explicit expressions for
them in the diagonal form.
The explicit form of $\alpha_i$ matrices (with brackets denoting
the diagonal elements) are,
$$
\alpha_1=e^{2t}, \alpha_2=(1,e^{3t}),\alpha_3=(e^{-2t}e^{t},e^{t}),
\alpha_4=
(e^{-t},e^{-t},e^{2t}),
$$
$$
\alpha_5=(e^{-3t},1,1), \alpha_6=(e^{-2t},e^t),\alpha_7= e^{-t}.  
$$
Knowledge of the explicit expressions for $\alpha$ matrices allow without 
any difficulties to resolve the first row of system (\ref{10}) and present
$r^1$ matrix is the following form:
$$ \def\quad{\hskip .45em}
r^1=\pmatrix{
e^{2t} & X & 0 & 0 & 0 & 0 & 0 & 0 & 0 & 0 & 0 & 0 & 0 & 0 & 0 \cr
     -X & 1 & 0 & A & 0 & 0 & 0 & 0 & 0 & 0 & 0 & 0 & 0 & 0 & 0 \cr
0 & 0 & e^{3t} & 0 & B & C & 0 & 0 & 0 & 0 & 0 & 0 & 0 & 0 & 0 \cr
0 & -A & 0 & e^{-2t} & 0 & 0 & 0 & 0 & 0 & 0 & 0 & 0 & 0 & 0 & 0 \cr
0 & 0 & -B & 0     & e^t & 0 & D & E & 0 & 0 & 0 & 0 & 0 & 0 & 0 \cr
0 & 0 & -C & 0     & 0 & e^t & F & G & 0 & 0 & 0 & 0 & 0 & 0 & 0 \cr
0 & 0 & 0 & 0     & -D & -F & e^{-t} & 0 & 0 & H & 0 & 0 & 0 & 0 & 0 \cr
0 & 0 & 0 & 0     & -E & -G & 0 & e^{-t} & 0 & K & 0 & 0 & 0 & 0 & 0 \cr
0 & 0 & 0 & 0     & 0 & 0 & 0 & 0 & e^{2t} & 0 & L & M & 0 & 0 & 0 \cr
0 & 0 & 0 & 0    & 0 & 0 & -H & -K & 0 & e^{-3t} & 0 & 0 & 0 & 0 & 0 \cr
0 & 0 & 0 & 0    & 0 & 0 & 0 & 0 &- L & 0       & 1 & 0 & N & 0 & 0 \cr
0 & 0 & 0 & 0    & 0 & 0 & 0 & 0 &- M & 0       & 0 & 1 & P & 0 & 0 \cr
0 & 0 & 0 & 0    & 0 & 0 & 0 & 0 & 0 & 0 & -N & -P & e^{-2t} & 0 & 0 \cr
0 & 0 & 0 & 0    & 0 & 0 & 0 & 0 & 0 & 0       & 0 & 0 & 0 & e^t & Y \cr
0 & 0 & 0 & 0    & 0 & 0 & 0 & 0 & 0 & 0       & 0 & 0 & 0 & -Y & e^{-t}
\cr}.
$$
The ansatz for $s^1$ matrix arise from the same for $r^1$, one by 
deleting diagonal elements and changing the remaining matrix to be
symmetrical
rather than antisymmetrical. In writing the above ansatz we have assumed 
additionally that $s^1$ may be chosen in symmetrical form. 
For the same reasons the ansatz for $r^2$ generator has the form:
$$ \def\quad{\hskip .5em}
r^2=\pmatrix{
e^t & 0 & y & 0 & 0 & 0 & 0 & 0 & 0 & 0 & 0 & 0 & 0 & 0 & 0 \cr
  0 & e^{2t} & 0 & 0 & n & p & 0 & 0 & 0 & 0 & 0 & 0 & 0 & 0 & 0 \cr
-y & 0 & e^{-t} & 0 & 0 & 0 & 0 & 0 & 0 & 0 & 0 & 0 & 0 & 0 & 0 \cr
0 & 0 & 0 & e^{3t} & 0 & 0 & h & k & 0 & 0 & 0 & 0 & 0 & 0 & 0 \cr
0 & -n & 0 & 0     & 1 & 0 & 0 & 0 & l & 0 & 0 & 0 & 0 & 0 & 0 \cr
0 & -p & 0 & 0     & 0 & 1 & 0 & 0 & m & 0 & 0 & 0 & 0 & 0 & 0 \cr
0 & 0 & 0 & -h     & 0 & 0 & e^t & 0 & 0 & 0 & d & f & 0 & 0 & 0 \cr
0 & 0 & 0 & -k     & 0 & 0 & 0 & e^t & 0 & 0 & e & g & 0 & 0 & 0 \cr
0 & 0 & 0 & 0     & -l & -m & 0 & 0 & e^{-2t} & 0 & 0 & 0 & 0 & 0 & 0 \cr
0 & 0 & 0 & 0    & 0 & 0 & 0 & 0 & 0 & e^{2t} & 0 & 0 & a & 0 & 0 \cr
0 & 0 & 0 & 0    & 0 & 0 & -d & -e & 0 & 0       & e^{-t} & 0 & 0 & b & 0
\cr
0 & 0 & 0 & 0    & 0 & 0 & -f & -g & 0 & 0       & 0 & e^{-t} & 0 & c & 0
\cr       
0 & 0 & 0 & 0    & 0 & 0 & 0 & 0 & 0 & -a & 0 & 0 & 1 & 0 & x \cr
0 & 0 & 0 & 0    & 0 & 0 & 0 & 0 & 0 & 0       & -b & -c & 0 & e^{-3t} & 0
\cr
0 & 0 & 0 & 0    & 0 & 0 & 0 & 0 & 0 & 0       & 0 & 0 & -x & 0 &e^{-2t}
\cr}.
$$
The anzates for $s^{1,2}$  obviously follows from the same for $r^{1,2}$.
In fact with the help of these ans\"atze we have resolved first two rows of 
equations of the system (\ref{14}).
We are now ready to solve the system (\ref{14}) as a whole. 
The first rows of second part of equations (\ref{14}) lead to the following 
values of $X,y$ ($a_{1,0}=b_{1,0}=0$):
$$
X^2=e^t \sinh t \sinh 2t,\quad y^2=\sinh^2 t
$$
Resolution of two next rows is the following:
\begin{eqnarray*}
& A^2=e^{-t}\sinh t\sinh 2t,\qquad B^2+C^2=e^{2t}\sinh t\sinh 3t, \\
& nB+pC=e^t yX, \qquad n^2+p^2=e^t\sinh t\sinh 2t.
\end{eqnarray*}
The last equations define two two-dimensional vectors with given angle
between them:
\begin{eqnarray*}
&(B,C)=e^t(\sinh t\sinh 3t)^{1\over 2}t(\cos \phi_1,\sin \phi_1), \\
&(n,p)=e^{{t\over 2}}(\sinh t\sinh 2t)^{1\over 2}(\cos \phi_2,\sin \phi_2),
\\
& \displaystyle \cos (\phi_1-\phi_2) = \left({\sinh t\over \sinh 3t}
\right)^{1\over 2}.
\end{eqnarray*}
This is analogue of the quantum angle mixed the states $\pi^0$ and $\Omega$ 
mesons in $(1,1)$ representation of $A_2$ algebra in the initial version
of $SU(3)$ symmetry of composite models of elementary particles.
In what follows we work in the gauge $\phi_2=0$ ($p=0$). But this choice of
gauge is absolutely inessential.
The system of equations of the third step is as follows:
\begin{eqnarray*}
&DF+EG=e^{-2t}BC=\sinh^2 t (2\cosh 2t)^{{1\over 2}}, \\
&D^2+E^2=e^{-2t}B^2+\sinh^2 t=2\sinh^2 t, \\
&F^2+G^2=e^{-2t}C^2+\sinh^2 t=\sinh t\sinh 3t,\\
&l^2=e^{-2t} n^2, \qquad m^2=e^{-2t} p^2=0, \qquad ml=e^{-2t} np=0, \\
&h^2+k^2=e^{2t} \sinh t \sinh 3t,\qquad m=0, \\
&Dh+Ek=e^t A n=e^t \sinh t\sinh 2t, \qquad Fh+Gk=e^t A p=0.
\end{eqnarray*}
The ans\"atze for $s^{1,2},r^{1,2}$ are obviously form invariant with
respect
to two-dimensional rotations in $(5-6),(7-8), (11-12)$ planes
(the reflection of this fact was the possibility to choose $p=0$ at
the second step
of the calculations with the help of rotations in (5,6) plane). Now we can
use this 
invariance choosing $E=0$ in the equations of the fourth step. After this
all 
equations above can be resolved without any difficulties with the result:
$$
D=\pm (2)^{{1\over 2}} \sinh t,\quad F=\pm\sinh t (\cosh 2t)^{{1\over 2}},
\quad G^2={1\over 2}\sinh^2 2t,
$$
$$
l^2=e^{-t} \sinh t\sinh 2t,\quad m=0,\quad h=\pm (2)^{-{1\over 2}} e^t
\sinh 2t,
\quad k=\pm e^t \sinh t (\cosh 2t)^{{1\over 2}}
$$
In the equations of the fifth step we use the $(11-12)$ invariance to fix
$M=0$. 
Keeping in mind this choice, we have:
\begin{eqnarray*}
&HK=e^{-2t}FG=\pm2^{-{1\over 2}}\sinh t \sinh 2t(\cosh 2t)^{{1\over 2}}, \\
&H^2=e^{-2t}(D^2+F^2-\sinh^2)={1\over 2}e^{-2t}\sinh^2 2t, \\
&K^2=e^{-2t}(G^2-\sinh^2 t)=e^{-2t} \sinh^2 t\cosh 2t, \\
&L^2=e^2 \sinh t\sinh 2t, \qquad dL+fM=e^t(Dl+Fm), \\
&eL+gM=e^t(El+Gm), \qquad d=D, \qquad e=E=0, \\
&e^2+g^2=e^{-2t}k^2+\sinh^2 t={1\over 2}\sinh^2 2t, \\
&de+fg=e^{-2t} hk=2^{-{1\over 2}}\sinh t \sinh 2t(\cosh 2t)^{{1\over 2}},
\\
&d^2+f^2=e^{-2t}h^2+\sinh^2 t=\sinh^2 t(\cosh 2t+2),\quad l^2=e^{-t}\sinh t
\sinh 2t.
\end{eqnarray*}
Solution of this system is trivial. We present below the final result:
$$
d=D,\quad e=E=0, \quad f=F,\quad g=G.
$$
The explicit form of $H$, $K$ is given above.
We will not reproduce here the two remaining steps of calculations because 
they do not contain anything new and to reconstruct them is a trivial
problem.

We note that always on the next following step we have 
the system of quadratic equations only for unknown matrix elements of 
$a_{n+1,n},a_{n,n+1},b_{n+1,n},b_{n,n+1}$ matrices, the right hand side of 
which was calculated on the previous step. On account of the invariance 
conditions the number of equations exactly equal  the number of unknown 
variables. The selfconsistency of the whole construction is only a
consequence
of the global representation theory.

\subsection{The general case of $(p,q)$ representation}  

To generalize  the results of the previous subsection to the case of 
arbitrary $(p,q)$ representation of $A_2$ algebra more detailed 
information about the structure of $a_{n,n\pm 1},b_{n,n\pm 1}$ is
necessary.
The first two rows of the system (\ref{14}) give additional restrictions on 
them. Namely different from zero are only those matrix elements 
$(a_{n,n\pm 1})_{s,s'},(b_{n,n\pm 1})_{s,s'}$ for which 
$(s,s')$ satisfy the conditions,
$$
{\alpha^s_n\over \alpha^{s'}_{n\pm 1}}=e^{\pm2t},\quad {\beta^s_n\over 
\beta^{s'}_{n\pm 1}}=e^{\pm2t}
$$
The diagonal elements of $\alpha,\beta$ matrices are in turn relating
by the relations,
$$
{\alpha^s_n\over \alpha^{s+1}_n}=e^{-3t},\quad {\beta^s_n\over 
\beta^{s+1}_n}=e^{3t}
$$
Let us denote the multiplicity of diagonal elements $\alpha^s_n,\beta^s_n$
as
$N^s_n$. Obviously $N_n=\sum_s N^s_n$.
The multiplicity $N^s_n$ is exactly equal to the natural number in the
corresponding
exponent in the Weyl character formula (\ref{WL}). After reducing this 
to the $A_1$ subgroup with the algebra of the first (second) simple root 
this is exactly the coefficient of $\alpha^s_n (\beta^s_n)$.

Each matrix $a_{n,n+1}$ (for definitness we choose the plus sign) 
is separated into $N^s_n\times N^{s'}_{n+1}$ rectangular block  matrices
$a^{s,s'}_{n,n+1}$,
where $(s,s')$ are the indices of diagonal matrix elements $\alpha^s_n,
\alpha^{s'}_{n+1}$ respectively. Only those
block matrices for which ${\alpha^s_n\over \alpha^{s'}_{n+1}}=e^{\pm2t}$
are nonzero.
Two important consequences follow from this fact. Firstly, on the line (
with
the ``wide" $N^s_n$) and on the column (with the "wide" $N^{s'}_{n+1}$) 
may be only one different from zero block matrix. Secondly,
all matrices with shifted on natural number indices $ a^{s\pm k, s'\pm
k}_{n,
n+1}$ (k - natural positive) are different from zero simultaneosly with
$a^{s,s'}_{n,n+1}$. Of course if the indices $s\pm k,s'\pm k$ are inside of
the domain of their definition.

{}From the explicit form of ans\"atze for $s^{1,2},r^{1,2}$ in the
beginning of 
this section follows their form invariance with respect to each
$SL(N^s_n;R)$
canonical transformations, which does not change diagonal matrix elements,
transforming $a$ matrices by the law:
$$
a\to G(N^s_n,R)aG(N^{s'}_{n+1})
$$
If we want to preserve symmetry of $s^{1,2}$ matrices (they are symmetrical 
by our convention up to now) it is necessary to reduce this transformation
up to
direct product of orthogonal $O_{N^s_n}$ transformations.
With similar considerations holding for the $b$ matrices, 
we are able to rewrite the remaining equations of the system (\ref{14}) 
(the three  last rows) in terms of only ``primitive'' block matrices:
$$
e^{-t} a^{s,s+2}_{n,n-1}a^{s+2,s}_{n-1,n}-e^t a^{s,s-2}_{n,n+1}a^{s-2,s}_
{n+1,n}=\alpha^s_n\sinh t \sinh (\ln \alpha^s_n) I_{N^s_n}
$$
\begin{equation}
e^{-{t\over 2}} b^{s,s+1}_{n,n+1}a^{s+1,s+3}_{n+1,n}=e^{{t\over 2}} 
a^{s,s+2}_{n,n-1}b^{s+2,s+3}_{n-1,n} \label{GGG}
\end{equation}
$$
e^{-t} bª{s,s-1}_{n,n-1} b^{s-1,s}_{n-1,n}-e^t b^{s,s+1}_{n,n+1} b^{s+1,s}_
{n+1,n}=-\sinh t (\lambda_n(\alpha^s_n)^{-1} \sinh (\ln
(\lambda_n(\alpha_n^s)^
{-1})) I_{N^s_n}
$$
where $I_{N^s_n}$ is the unity $N^s_n\times N^s_n$ matrix; the index $s$
runs
all values of $\alpha^s_n$.
The initial system (\ref{14}) is therefore split 
into a chain like system of equations for block matrices and each chain 
is living absolutely independent from the other ones.  
On each step of calculations we have to solve the following problem:
it is necessary to find the rectangular matrices of the given dimension
$a,b$, which satisfy the following system of algebraic equations:
\begin{equation}
a a^T=A,\quad b b^T=B,\quad b a^T=C,\quad (a b^t=C^T)\label{FL}
\end{equation}
where $A,B,C$ are the known matrices. In connection with (\ref{GGG}) they
are
defined on the previous step of calculation (exactly from the system
(\ref{FL}))
together with $\alpha^s_n,\beta^s_n$ known from the Weyl formula, which
 guarantees the  consistency of the whole construction.

Now we want to illustrate the consideration above by the the example of the 
previous subsection. The block structure of $s^{1,2},r^{1,2}$ ans\"atze is
the
following one:
$$
a^{2,0}_{1,2}=X,\quad a^{0,-2}_{2,3}=A,\quad a^{3,1}_{2,3}=\pmatrix{ B & C
\cr},
a^{1,-1}_{3,4}=\pmatrix{ D & E \cr
                         F & G \cr}
$$
$$
a^{-1,-3}_{4,5}=\pmatrix{ H \cr                         
                          K \cr},\quad a^{2,0}_{4,5}=\pmatrix{ L & M
                          \cr},\quad
a^{0,-2}_{5,6}=\pmatrix{ N \cr                         
                         P \cr},\quad a^{1,-1}_{6,7}=Y
$$
$$
b^{2,3}_{1,2}=y,\quad b^{0,1}_{2,3}=\pmatrix{ n & p \cr},\quad
b^{-2,-1}_{3,4}=\pmatrix{ h & k \cr},\quad b^{1,2}_{3,4}=\pmatrix{ l \cr
                                                                   m \cr}
$$
$$
b^{-1,0}_{4,5}=\pmatrix{ d & f \cr                         
                         e & g \cr},\quad b^{0,1}_{5,6}=\pmatrix{ b \cr
                                                                  c
                                                                \cr},\quad                                
b^{-3,-2}_{5,6}=a,\quad, b^{-2,-1}_{6,7}=x
$$
We would like to demonstrate the chain like structure with the 
example of equations relating primitive $a$ matrices:
\begin{eqnarray*}
&-e^ta_{1,2}^{2,0}a_{2,1}^{0,2}=-e^{2t}\sinh t\sinh 2t,\qquad
e^{-t}a_{2,1}^{0,2}a_{1,2}^{2,0}-e^ta_{2,3}^{0,-2}a_{3,2}^{-2,0}=0, \\
&e^{-t}a_{3,2}^{-2,0}a_{2,3}^{0,-2}=e^{-2t}\sinh t\sinh 2t, \qquad
-e^ta_{2,3}^{3,1}a_{3,2}^{1,3}=-e^{3t}\sinh t\sinh 3t, \\
&e^{-t}a_{3,2}^{1,3}a_{2,3}^{3,1}-e^{t}a_{3,4}^{1,-1}a_{4,3}^{-1,1} =
-e^t\sinh^2 t, \\
&e^{-t}a_{4,3}^{-1,1}a_{3,4}^{1,-1}-e^{t}a_{4,5}^{-1,-3}a_{5,4}^{-3,,-1} =
e^{-t}
\sinh^2 t, \\
&e^{-t}a_{5,4}^{-3,-1}a_{4,5}^{-1,-3}=e^{-3t}\sinh t\sinh 2t, \qquad
-e^ta_{4,5}^{2,0}a_{5,4}^{0,2}=-e^{2t}\sinh t\sinh 2t, \\ 
&e^{-t}a_{5,4}^{0,2}a_{4,5}^{2,0}-e^ta_{5,6}^{0,-2}a_{5,6}^{-2,0}=0, \qquad
e^{t}a_{5,6}^{-2,0}a_{6,5}^{0,-2}=e^{2t}\sinh t\sinh 2t, \\
&e^ta_{6,7}^{1,-1}a_{7,6}^{-1,1}=e^t\sinh^2 t.
\end{eqnarray*}
Here we have two chains with 3 elements (spin 1), one chain
with four elements (spin ${3\over 2}$) and one scalar chain. 

\sect{The case of $B_2$ algebra}

In this case $p=2$. We work again in representation with  diagonal $Q$. 
{}From (\ref{10}) it   follows that its diagonal matrix elements 
are relating by the condition ${\lambda_i\over \lambda_{i+1}}=e^{2t}$. 
It follows also that generators $Q^{\pm}_1$  commute with $Q$ and so 
 $s^1,r^1$ have the form:
$$
s^1=\pmatrix{...& a_i &....\cr},\quad r^1=\pmatrix{...& b_i &...\cr}
$$
 $s^2,r^2$ maintain their  previous  forms, 
$$
s^2=\pmatrix{....& a_{i,i-1} & 0 & a_{i,i+1}... \cr}\quad
r^2=\pmatrix{....& -a_{i,i-1} & \alpha_i & a_{i,i+1}... \cr}
$$
The first equation (\ref{5}) preserves its form for each component of 
the ansatz for $(s^1,r^1)$ matrices:
\begin{equation}
[a_i,b_i]=\tanh t(a_i^2-b_i^2+1)\label{16}
\end{equation}

The second equation (\ref{5}) is equivalent to (\ref{11}) with the
replacement 
$t\to 2t$:
$$
e^{-2t} a_{n,n-1} \alpha_{n-1}=e^{2t} \alpha_n a_{n,n-1},\quad 
e^{2t} a_{n,n+1} \alpha_{n+1}=e^{-2t} \alpha_n a_{n,n+1}, 
$$
\begin{equation}
2e^{-2t} a_{n,n-1}a_{n-1,n}-2e^{2t} a_{n,n+1}a_{n+1,n}=\sinh
t(I_n-\alpha_n^2)
\label{17}
\end{equation}

The system of equations of mixed unknown functions in the $(s^1,r^1)$ and
$(s^2,r^2)$ ans\"atze has the form:
$$
e^{-t}(a_i+b_i)a_{i,i-1}=e^t a_{i,i-1} (a_{i-1}+b_{i-1})\quad  
e^t(a_i-b_i)a_{i,i+1}=e^{-t} a_{i,i+1} (a_{i+1}-b_{i+1})
$$
\begin{equation}
e^t \alpha_i (a_i+b_i)-e^{-t}(a_i+b_i) \alpha_i=2\sinh t\lambda_i I_i,\quad
e^{-t} \alpha_i (a_i-b_i)-e^t (a_i-b_i) \alpha_i=2\sinh t\lambda_i I_i 
\label{18}
\end{equation}

The Weyl group of $B_2$ algebra consists of 8 elements. The character of
its
$(p,q)$ representation calculated with the help of (\ref{WL}) has the form:
\begin{equation}
\pi^{(p,q)}(\tau_1,\tau_2)={\sinh l_1\tilde \tau_2\sinh l_2\tilde \tau_2-
\sinh l_1\tilde \tau_2\sinh l_2\tilde \tau_1\over \sinh 2\tilde \tau_1\sinh 
\tilde \tau_2-\sinh \tilde \tau_1\sinh 2\tilde \tau_2}\label{WB}
\end{equation}
where $l_1-l_2=p+1,l_2=q+1,l_1=p+q+2,l_1+l_2=p+q+3,\tilde \tau_1=\tau_1,
\tilde \tau_2=\tau_2-\tau_1$. ( Really (\ref{WB}) is the character of $C_2$ 
algebra).

All diagonal elements of the matrices
$R_1,R_2,Q=R_1R_2$ necessary for further calculations  
have to be obtained from (\ref{WB}).
Now we would like to consider in details the concrete case of the $(1,1)$ 
16  dimensional representation of $B_2$ algebra.

\subsection{The case of $(1,1)$ representation}

In this case $l_1=4,l_2=2$ and the character of the $(1,1)$ representation
in 
correspondence with (\ref{WB}) has the form:
$$ 
\pi^{(1,1)}(\tau_1,\tau_2)=(e^{\tilde \tau_1}+e^{-\tilde \tau_1})
(e^{\tilde \tau_2}+e^{-\tilde \tau 2})(e^{\tilde \tau_1}+e^{\tilde \tau_2})
(1+e^{-(\tilde \tau_1+\tilde \tau_2)})=
$$
$$
e^{2\tau_2-\tau_1}+e^{\tau_2+\tau_1}+e^{2\tau_2-3\tau_1}+2e^{\tau_2-\tau_1}
+
2e^{\tau_1}+e^{-\tau_2+3\tau_1}+
$$
$$
e^{\tau_2-3\tau_1}+2e^{-\tau_1}+2e^{-\tau_2+\tau_1}+e^{-2\tau_2+3\tau_1}+
e^{-\tau_2-\tau_1}+e^{-2\tau_2+\tau_1}
$$
The order of the exponents in the latter expression is chosen so as to 
present the final expressions in more attractive form.

After reducing to the $A_1$ subgroup with the algebra of the first 
complicate root of $B_2$ algebra ($X^{\pm}_{1,2},H=h_1+2h_2$), what is 
equivalent to substituting $\tau_1=t,\tau_2=2t$ into the expression for 
the character, we obtain,
$$
\pi^{(1,1)}(t,2t)=2e^{3t}+6e^{t}+6e^{-t}+2e^{-3t}
$$
Reducing to the $A_1$ subgroup of the first simple root ($X^{\pm}_1,H=h_1,
\tau_1=t,\tau_2=0$) leads to:
$$
\pi^{(1,1)}(t,0)=e^{-t}+e^{t}+e^{-3t}+2e^{-t}+2e^{t}+e^{3t}+e^{-3t}+2e^{-t}
+
2e^{t}+e^{3t}+e^{-t}+e^{t}
$$
Using these details we obtain the explicit form of the $\alpha$ matrices:
$$
\alpha_1=(e^{4t},e^{2t}),\quad \alpha_2=(e^{4t},e^{2t},e^{2t},1,1,e^{-2t}),
$$
$$
\alpha_3=(e^{2t},1,1,e^{-2t},e^{-2t},e^{-4t}),\quad
\alpha_4=(e^{-2t},e^{-4t})
$$
(as above the brackets contain the diagonal elements of the $\alpha$
matrices).

The rectangular matrices from the ($s^2,r^2$) ans\"atze have the following
dimension:
$$
a_{1,2}\to 2\times 6,\quad a_{2,3}\to 6\times 6,\quad a_{3,4}\to 6\times 2.
$$
and the $a_i,b_i$ matrices from the ($s^1,r^1$) ans\"atze,
$$
a_1,b_1\to 2\times 2,\quad a_2,b_2\to 6\times 6,\quad a_3,b_3\to 6\times 6,
\quad a_4,b_4\to 2\times 2
$$

The first row equations of the system (\ref{17}) has the following solution
\begin{eqnarray*}
&a_{1,2}=\pmatrix{0 & 0 & 0 & A & B & 0 \cr
                 0 & 0 & 0 & 0 & 0 & C \cr},\qquad
a_{2,3}=\pmatrix{0 & D & E & 0 & 0 & 0 \cr                 
                 0 & 0 & 0 & F & G & 0 \cr
                 0 & 0 & 0 & H & K & 0 \cr
                 0 & 0 & 0 & 0 & 0 & L \cr
                 0 & 0 & 0 & 0 & 0 & M \cr
                 0 & 0 & 0 & 0 & 0 & 0 \cr}, \\
&a_{3,4}=\pmatrix{N & 0 \cr                 
                 0 & P \cr
                 0 & R \cr
                 0 & 0 \cr
                 0 & 0 \cr
                 0 & 0 \cr}
\end{eqnarray*}                  
Or in the notations of primitive rectangular matrices, introduced in the 
previous section:                  
$$
a_{1,2}^{2,-2}=C,\quad a_{1,2}^{4,0}=\pmatrix{A & B \cr},\quad 
a_{2,3}^{4,0}=\pmatrix{D & E \cr},\quad a_{2,3}^{2,-2}=\pmatrix{F & G \cr
                                                                H & K \cr},
$$
$$
a_{2,3}^{0,-4}=\pmatrix{L \cr                   
                        M \cr},
a_{3,4}^{0,-4}=\pmatrix{P \cr                   
                        R \cr}, a_{3,4}^{2,-2}=N
$$
We draw attention to the fact that the notations in terms of primitive 
rectangular matrices are independent of the choice of the order 
of exponents in the Weyl character formula.

The second row of the system (\ref{17}) is as follows:
\begin{eqnarray*}
&-e^{2t} a_{1,2}^{2,-2}a_{2,1}^{-2,2}=-e^{2t}\sinh^2 2t, \qquad
-e^{2t} a_{1,2}^{4,0}a_{2,1}^{0,4}=-e^{4t}\sinh 2t \sinh 4t, \\
&e^{-2t} a_{2,1}^{-2,2}a_{1,2}^{2,-2}=e^{-2t}\sinh^2 2t, \qquad
e^{-2t} a_{2,1}^{0,4}a_{1,2}^{4,0}-e^{2t} a_{2,3}^{0,-4}a_{3,2}^{-4,0}=0,
\\
&-e^{2t} a_{2,3}^{2,-2}a_{3,2}^{-2,2}=-e^{2t}\sinh^2 2t I_2, \qquad
e^{-2t} a_{3,2}^{-4,0}a_{2,3}^{0,-4}=e^{-4t}\sinh 2t\sinh 4t, \\
&e^{-2t} a_{3,2}^{-2,2}a_{2,3}^{2,-2}=e^{-2t}\sinh^2 2t I_2, \qquad
e^{-2t} a_{3,2}^{0,4}a_{2,3}^{4,0}-e^{2t} a_{3,4}^{0,-4}a_{4,3}^{-4,0}=0,
\\
&e^{-2t} a_{4,3}^{-4,0}a_{3,4}^{0,-4}=e^{-4t}\sinh 2t\sinh 4t, \qquad
e^{-2t} a_{4,3}^{-2,2}a_{3,4}^{2,-2}=e^{-2t}\sinh^2 2t.
\end{eqnarray*}
Not all of the equations above are different but no contradictions between
them
is observed, what can serve as one additional argument of selfconsistency
of 
the whole construction. We present below only nonrepeated equations:
\begin{eqnarray*}
&C^2=\sinh^2 2t,\quad A^2+B^2=e^{2t}\sinh 2t\sinh 4t,\quad, L=\pm
e^{-2t}A,\quad
M=\pm e^{-2t}B, \\
& \displaystyle \pmatrix{F & G \cr H & K \cr}=
\sinh 2t \pmatrix{\cos \phi & \sin \phi \cr
         -\sin \phi & \cos \phi \cr}, \\                  
&D=\pm e^{2t}P,\quad E=\pm e^{2t}R,\quad P^2+R^2=e^{-2t}\sinh 2t \sinh
4t,\quad
N^2=\sinh^2 2t
\end{eqnarray*}
{}From these we conclude that their solution may be obtained uniquely
up to orthogonal transformations.                  
                  
Now we pass to equations of the second row of (\ref{18}) relating $\alpha$
and 
$a\pm b$ matrices. Direct calculations lead to the result:
$$
a_1+b_1=\pmatrix{e^{-t} & 0 \cr                  
                      x & e^t \cr},\quad
a_1-b_1=\pmatrix{-e^{-t} & y \cr 
                      0 & -e^t \cr}
$$            
$$
a_4+b_4=\pmatrix{e^{-t} & 0 \cr                  
                      u & e^t \cr},\quad
a_4-b_4=\pmatrix{-e^{-t} & v \cr 
                      0 & -e^t \cr}
$$
$$
a_2+b_2=\pmatrix{e^{-3t} & 0 & 0 & 0 \cr
            X_{2,1} & e^{-t} & 0 & 0\cr                
                  0 & X_{3,2} & e^t & 0 \cr
                  0 & 0 & X_{4,3} & e^{3t} \cr},\quad
a_2-b_2=\pmatrix{-e^{-3t} & Y_{1,2} & 0 & 0 \cr
                       0 & -e^{-t} & Y_{2,3} & 0\cr                
                       0 & 0 & -e^t & Y_{3,4} \cr
                       0 & 0 & 0 & -e^{3t} \cr}
$$
$$                                        
a_3+b_3=\pmatrix{e^{-3t} & 0 & 0 & 0 \cr
                 U_{2,1} & e^{-t} & 0 & 0\cr                
                       0 & U_{3,2} & e^t & 0 \cr
                       0 & 0 & U_{4,3} & e^{3t} \cr},\quad
a_3-b_3=\pmatrix{-e^{-3t} & V_{1,2} & 0 & 0 \cr
                       0 & -e^{-t} & V_{2,3} & 0\cr                
                       0 & 0 & -e^t & V_{3,4} \cr
                       0 & 0 & 0 & -e^{3t} \cr}
$$
We have presented the $6\times 6$ matrices $a_2\pm b_2,a_3\pm b_3$ in 
four-dimensional form, keeping in mind that in these expressions $2,3$
indices indeed are two-dimensional ones. Thus $X_{21}$ and $Y_{12}$ are
$2\times 1$ and $1\times 2$ two-dimensional column and row vectors
respectively
and so on.

Now it is necessary to take into account that $a,b$ matrices satisfy  
equations (\ref{16}). In terms of $(a\pm b)$ matrices the last equations
may be 
rewritten as:
$$
e^t(a+b)(b-a)-e^{-t}(b-a)(a+b)=2\sinh t I
$$
Direct substitution of expressions for $(a_1,b_1),(a_4,b_4)$ obtained above 
into the last equation yields the following restriction on 
parameters:
$$
xy=4\sinh^2 t,\quad uv=4\sinh^2 t
$$
If we want a symmetrical $s^1$ matrix, then we obtain
$$
x=y=u=v=2\sinh t
$$
This result is possible to understand without any calculations by
consideration of the relations of the section 3, 
${a+b\over 2\sinh t}=Q^+,{a-b\over 2\sinh t}=Q^-$. 
The expressions for $(a_{1,4}\pm b_{1,4}$ coincide with those for
spinor (two-dimensional) representation of quantum $A^q_1$ algebra with 
additional similarity transformation with the two-dimensional matrix
$\sigma=
\pmatrix{0 & 1 \cr                      
         1 & 0 \cr}$.
The situation with the six-dimensional matrices is the same. It is
necessary
to consider direct sum of four-dimensional (${3\over 2}$ spin)
representation
of $A^{q}_1$, related to 1,2,5,6 indices and two dimensional (${1\over 2}$ 
spin) one, related to the 3,4 indices. After this do similarity 
transformation with $6\times 6$ matrix with nonzero unities on
the main antidiagonal. Result will be exactly matrices $a_{2,3}\pm
b_{2,3}$.
Under such a procedure the following limitations on parameters $X,Y,U,V$
arise:
\begin{eqnarray*}
& \displaystyle X_{2,1}=U_{2,1}=\pmatrix{2e^{-t}(\sinh 3t\sinh t)^{{1\over
2}} \cr
                                                                     0 \cr}
,\quad X_{2,3}=U_{2,3}=\pmatrix{0 & 2\sinh t\cr
                        2\sinh 2t & 0 \cr}, \\
& \displaystyle X_{3,4}=U_{3,4}=\pmatrix{0 & 2e^{t}(\sinh 3t\sinh
t)^{{1\over 2}} \cr}
\end{eqnarray*}
and analogous expressions for $Y,V$ elements ($Y=X^T,V=U^T$). It is 
necessary to emphasise that  we have fixed the gauge and thus
all expressions above do not contain any additional parameters. 
But this was achieved by the definite choice of the form of the
$x,y,X,Y,u,v,U,V$  matrices.

At last we have to satisfy the first row of equations (\ref{18}) relating 
rectangular matrices $a_{n,n+1}$ to the  matrices $a_i\pm b_i$.
Direct calculation leads to the following final result:
$$
e^{-t}(X_{4,3} a^{0,4}_{2,1})=e^tCx,\quad e^{-t}(U_{3,2} a^{4,0}_{2,3})=
e^t (a^{2,-2}_{23}X_{2,1}),\quad 
$$
$$
e^{-t}(U_{4,3} a^{2,-2}_{23})=e^t(a^{-4,0}_{3,2}X_{32}),\quad
e^{-t}u N=e^t(a^{-4,0}_{4,3} U_{2,1})
$$
These uniquely fixed all parameters in the construction,
$$
A=E=e^t\sinh 2t ({\sinh 5t\over \sinh 3t})^{1\over 2}\quad
B=D=e^t\sinh 2t ({\sinh t\over \sinh 3t})^{1\over 2},
$$
$$
\sin \phi=-{\sinh 2t\over \sinh 3t},\quad \cos \phi=={(\sinh 5t \sinh t)^
{1\over 2}\over \sinh 3t}
$$                                                                         
                  
\section{The case of $G_2$ algebra}

Repeating  the arguments of two previous sections (the cases of
$A_2$ and $B_2$ algebras) we write down to the following ans\"atze for 
the matrices $(s^1,r^1)$, $(s^2,r^2)$, which we present in symbolical row 
form:
$$ \def\quad{\hskip .5em}
s^1=\pmatrix{....& a_{i,i-1} & 0 & a_{i,i+1}... \cr},\qquad
r^1=\pmatrix{....& a_{i,i-1} & \alpha_i & -a_{i,i+1}... \cr}
$$
$$ \def\quad{\hskip .5em}
s^2=\pmatrix{....& b_{i,i-3} & 0 & 0 & 0 & 0 & 0 & b_{i,i+3}... \cr},\qquad
r^2=\pmatrix{....& -b_{i,i-3} & 0 & 0 & \beta_i & 0 & 0 & b_{i,i+3}... \cr}
$$
The sign differences, $a_{i,i+1}$ instead of $a_{i,i-1}$ in $A_2$ algebra 
case, are related to the different signs in the first equation
(\ref{8}). In these formulae the notation is the same as in the 
previous sections: $\alpha_i,\beta_i$ are quadratic $N_i\times N_i$
matrices,
$a_{p,q},b_{p,q}$ are $N_p\times N_q$ rectangular ones.
The equation relating $s^1,r^1$ is equivalent to a matrix system:
$$
e^{t} a_{n,n-1} \alpha_{n-1}=e^{-t} \alpha_n a_{n,n-1},\quad 
e^{-t} a_{n,n+1} \alpha_{n+1}=e^{t} \alpha_n a_{n,n+1}, 
$$
\begin{equation}
-2e^t a_{n,n-1}a_{n-1,n}+2e^{-t} a_{n,n+1}a_{n+1,n}=\sinh t(I_n-\alpha_n^2)
\label{G1}
\end{equation}
and that for $s^2,r^2$ leads to:
$$
e^{-3t} b_{n,n-3} \beta_{n-3}=e^{3t} \beta_n b_{n,n-3},\quad 
e^{3t} b_{n,n+3} \beta_{n+3}=e^{-3t} \beta_n b_{n,n+3}, 
$$
\begin{equation}
2e^{-3t} b_{n,n-3} b_{n-3,n}-2e^{3t} b_{n,n+3} b_{n+3,n}=\sinh
t(I_n-\beta_n^2)
\label{G2}
\end{equation}
where $I_n$ is quadratic unity $N_n\times N_n$ matrix.
The last equation (\ref{6}) and definition of $Q$  (\ref{6A})
imply the system for mixed $a,\alpha$ and $b,\beta$ matrices:
\begin{equation}
\begin{array}{rll}
& e^{{3t\over 2}} b_{n,n-3}\alpha_{n-3}\ =\ 
e^{-{3t\over 2}}\alpha_n b_{n,n-3}\quad ,\quad 
e^{{3t\over 2}} a_{n,n+1}\beta_{n+1}=e^{-{3t\over 2}}\beta_n a_{n,n+1}\ ,\\ 
&
e^{-{3t\over 2}} a_{n,n-1}\beta_{n-1}=e^{{3t\over 2}}\beta_n a_{n,n-1}
\quad ,\quad 
e^{-{3t\over 2}} b_{n,n+3}\beta_{n+3}=e^{{3t\over 2}}\alpha_n b_{n,n+3}\
,\\
&
e^{-{3t\over 2}}b_{n,n+3}a_{n+3,n+4}=e^{{3t\over 2}}a_{n,n+1}b_{n+1,n+4}
\quad ,\quad
e^{{3t\over 2}}b_{n,n-3}a_{n-3,n-4}=e^{-{3t\over 2}}a_{n,n-1}b_{n-1,n-4}\
,\\
&
\alpha_n \beta_n=\lambda_n I_n
\end{array}\label{G3}
\end{equation}
The last equation allows the elimination of matrices $\beta_n$
to rewrite (\ref{G1}), (\ref{G2}),(\ref{G3}) in more compact form. We
assume also that matrices $s^1, s^2$ are symmetrical in that,
$$
a_{i+1,i}=a^T_{i,i+1},\quad b_{i+3,i}=b^T_{i,i+3}\ .
$$
Finally the system of equations to be solved in the case of $G_2$ algebra
takes 
the form:
\begin{equation}
\begin{array}{rll}
&& e^{t} a_{n,n-1} \alpha_{n-1}=e^{-t} \alpha_n a_{n,n-1},\quad 
e^{-t} a_{n,n+1} \alpha_{n+1}=e^{t} \alpha_n a_{n,n+1}, \\
&&
e^{{3t\over 2}} b_{n,n-3}\alpha_{n-3}=e^{-{3t\over 2}}\alpha_n
b_{n,n-3},\quad 
e^{-{3t\over 2}} b_{n,n+3}\alpha_{n+3}=e^{{3t\over 2}}\alpha_n b_{n,n+3}, 
\\ &&
-2e^t a_{n,n-1}a_{n-1,n}+2e^{-t} a_{n,n+1}a_{n+1,n}=\sinh t(I_n-\alpha_n^2)
\\ &&
2e^{3t} b_{n,n-3} b_{n-3,n}-2e^{-3t} b_{n,n+3} b_{n+3,n}=\sinh
t(I_n-\beta_n^2)
\\ &&
e^{{3t\over 2}} b_{n,n-3}a_{n-3,n-4}=e^{-{3t\over 2}} a_{n,n-1}b_{n-1,n-4}
\label{GG}
\end{array}
\end{equation}
Here, as in the previous sections, the diagonal matrices 
$\alpha_n,\beta_n$ are considered to be known from the Weyl formula
(\ref{WL});
unknown are the rectangular matrices $a,b$ of the corresponding dimension.

As in the case of $A_2$ algebra rectangular matrices $a,b$ have  block 
structure in terms of primitive component of which the system (\ref{GG})
may be
presented. We omit this general consideration restricting ourselves
to the simplest example of the first fundamental 
representation of quantum $G_2$ algebra only, with the aim of demonstrating
the consistency of the proposed construction.

\subsection{ The case of $(1,0)$ representation}

necessary 
in
following
The standard basis, using lowering operators, of the $(1,0)$ representation 
is given by
$$
X^-_1\ve{1}, \ve{1}, X^-_1 X^-_1X^-_2 X^-_1\ve{1}, X^-_1X^-_2 X^-_1\ve{1},
$$
$$
X^-_2 X^-_1\ve{1}, X^-_1X^-_2X^-_1 X^-_1X^-_2 X^-_1\ve{1}, X^-_2 X^-_1
X^-_1X^-_2 X^-_1\ve{1}
$$
On this basis the generator $Q=\exp (h_1+3h_2)t$ takes the values
$$
Q=(e^{2t},e^t,e^t,1,e^{-t},e^{-t},e^{-2t})
$$
and $R_{1,2}$,
$$
R_1\equiv \exp h_1t=(e^{-t},e^t,e^{-2t},1,e^{2t},e^{-t},e^t)
$$
$$
R_2\equiv \exp 3h_2t=(e^{3t},1,e^{3t},1,e^{-3t},1,e^{-3t})
$$
The explicit expressions for the $\alpha$ matrices are as follows:
$$
\alpha_1=e^{-t}\ ,\ \alpha_2=\pmatrix{ e^t & 0 \cr
                                    0 & e^{-2t} \cr},\ 
\alpha_3=1\ ,\ 
\alpha_4=\pmatrix{ e^{2t} & 0 \cr
               0 & e^{-t} \cr},\  
\alpha_5=e^t
$$
According to the general scheme nonzero are the 
following nondiagonal elements of $(s^1,r^1)$ matrices:
$$
a_{1,2}=(A,B),\quad a_{3,2}=(C,D),\quad a_{3,4}=(E,F),\quad a_{5,4}=(G,H)
$$
and elements of $(s^2,r^2)$ ones:
$$
b_{1,4}=(x,y),\quad b_{5,2}=(u,v)
$$
where all the parameters are unknown and need to be determined 
from the system (\ref{GG}).
After substitution into equations of the first row of the system (\ref{GG})
using the explicit expressions for $\alpha$ matrices we obtain:
$$
a_{1,2}=(A,0),\quad a_{3,2}=(0,D),\quad a_{3,4}=(E,0),\quad a_{5,4}=(0,H)
$$
$$
b_{1,4}=(x,0),\quad b_{5,2}=(0,v)
$$
{}From the next two rows of the system (\ref{GG}) it uniquely follows that
\begin{eqnarray*}
&A^2=\sinh^2 t, \qquad D^2=e^{-t}\sinh t \sinh 2t, \\
&E^2=e^t\sinh t\sinh 2t, \qquad H^2=\sinh^2 t, \qquad x^2=v^2=\sinh^2 3t.
\end{eqnarray*}
By the direct check one can verify that equations of the last row of the
system (\ref{GG}) are satisfied automatically.

\section{Outlook}

The results of the present paper are simultaneously absolutely unexpected
and surprising at least to the author. Long time peoples were sure that
the solution of the problem of explicit form of infinitesimal generators
for
an arbitrary representation must be in a deep connection with the bases
defined 
by the  eigenvalues of the necessary number of mutually commuting operators 
constructed from the generators of the corresponding algebra. The values of 
Casimir operators in such construction define the indices of the
irreducible 
representation. 
The solution of the problem in this way was found in the famous papers
of I.~M.~Gelfand and M.~L.~Tsetlin fifty years ago \cite{G1}. These 
authors were able to represent in explicit form the infinitesimal operators
for an arbitrary representation of classical semisimple series $A_n$,
$B_n$,
and $D_n$. However all other numerous attempts to generalize these result
to all
remaining semisimple series were unsuccessful.

The way proposed in the present paper  deals neither with the Casimir 
operators nor with a family of mutually commutating operators and their 
eigenvalues. Only knowledge of the result of the action of the group 
operator $e^\tau$ on the basis state vectors is essential and this yields a 
possibility to solve the problem also for the case of quantum algebras.

The case of the usual semisimple algebras arises in the limit of the 
deformation parameter going to zero. 
Usually the back way of the thought is used:
it is necessary firstly to construct a representation of a semisimple
algebra 
and then generalize it to the quantum algebra case.

The algorithm of the present paper is similar to a computer program in that
it is necessary to iterate the same operation: to solve the 
algebraical system of equations the inhomogeneous part of which is known
from 
the previous steps of calculations.

Of course the key point of the whole construction is the Weyl formula for 
the characters of  finite-dimensional irreducible representations of  
semisimple groups. It guarantees the selfconsistency of the whole
construction 
and gives the necessary number of initial parameters (the explicit values
of 
diagonal elements of $\alpha,\beta$ matrices and their multiplicities)
through 
which the generators of the simple roots are expressed.
The invariant character of the Weyl formula allows the hope that the
problem of 
explicit realisation of the generators of the simple roots for quantum 
algebras may also be solved in invariant terms. Unfortunately at this
moment 
these terms are unknown to the author.

\noindent{\bf Acknowledgements.}

The author would like to thank C. Devchand, A.V. Razumov and Nodari
Vakhania 
for fruitful discussions and big help in preparation of the manuscript for 
publication. This paper was done under the partial support of the 
Russian Foundation for Fundamental Researches (RFFI) GRANT--N 98-01-00330.

\end{document}